\documentclass[twocolumn,showpacs,preprintnumbers,amsmath,amssymb]{revtex4}
\usepackage{graphicx}
\usepackage{dcolumn}
\usepackage{bm}
\usepackage{bbold}
\usepackage{booktabs}

\begin{document}

\preprint{}

\title{Low-energy dynamics of the two-dimensional \\
S=1/2 Heisenberg antiferromagnet on percolating clusters}

\author{Ling~Wang}
 \affiliation{Department of Physics, Boston University, 590 Commonwealth Avenue,
Boston, Massechussett, 02215}
\author{Anders~W.~Sandvik}
 \affiliation{Department of Physics, Boston University, 590 Commonwealth Avenue,
Boston, Massechussett, 02215}

\date{June 8, 2006}

\begin{abstract}
  We investigate the quantum dynamics of site diluted $S=1/2$
  Heisenberg antiferromagnetic clusters at the percolation threshold.
  We use Lanczos diagonalization to calculate the lowest excitation gap $\Delta$ 
  and, to reach larger sizes, study an upper bound for $\Delta$ 
  obtained from sum rules involving 
  the staggered structure factor and susceptibility, which we evaluate by quantum 
  Monte Carlo simulations. Scaling the gap distribution with the cluster length 
  $L$, $\Delta \sim L^{-z}$, we obtain a dynamic exponent $z \approx 2D_f$, where 
  $D_f=91/48$ is the fractal dimensionality of the percolating cluster. This is in 
  contrast to previous expectations of $z=D_f$. We argue that the low-energy
  excitations are due to weakly coupled effective moments formed due to
  local imbalance in sublattice occupation.
\end{abstract}

                             
\maketitle

Antiferromagnets under doping with spinless vacancies exhibit
interesting phenomena resulting from quantum fluctuations building on
top of fluctuations of classical percolation. Various calculations for
the two-dimensional (2D) Heisenberg model indicated that the quantum
fluctuations for $S=1/2$ are strong enough to cause a quantum phase
transition at some vacancy concentration $p=p_c$ less than the
classical percolation threshold $p^*$ \cite{qtransition}.
However, recent quantum Monte Carlo (QMC) simulations have
demonstrated that long-range order survives for all $p \le p^*$
\cite{kato00,andersPRB2002}. Experiments on magnetically diluted
layered cuprates agree with this, showing a
low-temperature divergent correlation length up to the percolation
point \cite{vajkScience2002}.  The fractal percolating cluster at
$p^*$ is ordered \cite{andersPRB2002}. The dilution-driven phase
transition is therefore a classical percolation transition of
the static spin order, but quantum fluctuations present in the 
percolating cluster lead to changes in quantities related to the low-energy 
dynamics. The critical exponents depend on classical percolation exponents 
and the dynamic exponent $z$ of the clusters \cite{vojtaPRL2005}, which 
thus becomes the focus of attention---it is the subject of this Letter.

The singlet-triplet gap of a clean $D$-dimensional antiferromagnet on a lattice 
of even length $L$ scales as $L^{-z}$ with $z=D$ \cite{hasenfratz}. The lowest 
excitations are the quantum rotor states responsible for symmetry breaking  
when $L \to \infty$. Vojta and Schmalian recently generalized this result to 
percolating clusters; $z=D_f$ \cite{vojtaPRL2005}, where in 2D the fractal
dimensionality $D_f=91/48$ \cite{Stauffer}. 
QMC calculations by Yu {\it et al.}~\cite{yuPRL05} 
seem to confirm $z=D_f$ at $p^*$, however in an indirect way from the 
finite-temperature scaling $\xi \sim T^{-1/z}$ of the correlation length. This 
form is expected for a quantum-critical system \cite{chn88} but is not necessarily 
related to the finite-size gap of ordered clusters \cite{note}. The 
exponent governing the finite-size gap is the one relevant to the 
$T=0$ dilution transition. Here we will take a more direct approach, 
studying the scaling of the excitation gaps of percolating
clusters. Our main result of this study is that the dynamic exponent
is much larger than expected; $z \approx 2D_f$. We will argue that the
lowest excitations are due to weak interactions between moments  formed 
by local sublattice asymmetry.

The diluted Heisenberg hamiltonian is
\begin{equation}
\label{hamiltonian}
  H=J\sum_{\langle i,j\rangle}\epsilon_i \epsilon_j 
{\bf S}_i\cdot {\bf S}_j,\quad (J>0),
\end{equation}
where $\langle i,j\rangle $ are nearest neighbors on a 2D square
lattice and $\epsilon_i=0,1$ with probability $p$ and $1-p$. 
We focus on the percolation threshold, $p=p^* \approx 0.407$
\cite{Stauffer} and study the properties of clusters grown on 
an infinite lattice, keeping clusters of a target size $n$ 
(see Ref.~\cite{andersPRB2002}). We first consider the gap
$\Delta = E_1 - E_0$ between the singlet ground state and triplet
first excited state of clusters with an equal number of spins in the
two ($A,B$) sublattices; $n=n_A+n_B$, $n_A=n_B$.  We also discuss $n_A \not= n_B$, 
for which the ground state is an $S=|n_A - n_B|/2$ multiplet, and find a
subtle difference in the nature of the low-energy excitations.

Using Lanczos diagonalization \cite{TRLan}, we study clusters with $n$ up to $20$.
In order to reach larger sizes we use a QMC method (stochastic series
expansion, SSE \cite{andersPRB1999}).  Due to statistical errors, it is
not possible to evaluate very small gaps directly using QMC.  We therefore
consider an upper bound $\Delta^*$ for the gap,
\begin{equation}
\Delta^* = 2S(\pi,\pi)/\chi(\pi,\pi) \ge \Delta,
\label{gapbound}
\end{equation}
obtained from well known sum rules for the static structure factor 
$S({\bf q})$ and susceptibility $\chi({\bf q})$;
\begin{eqnarray}  
  \int_0^{\infty}d\omega S({\bf q},\omega)&=&S({\bf q}),\\
  2\int_0^{\infty}\frac{d\omega}{\omega}S({\bf q},\omega)&=&\chi({\bf q}),
  \label{sumrules}
\end{eqnarray}
where $S({\bf q},\omega)$ is the dynamic structure factor. 
For a single mode $\Delta^*=\Delta$.  As in the undiluted system \cite{hasenfratz},
we expect $\Delta^*$ for Heisenberg clusters to scale as the true gap; 
$\Delta^* \sim L^{-z} \sim n^{-z/D_f}$. At the very least, $z$ obtained from 
$\Delta^*$ will be a {\it lower bound} for the dynamic exponent.

In the SSE simulations we reach the ground state using a $\beta = J/T$
doubling procedure \cite{andersPRB2002}, checking all clusters
individually for $T \to 0$ saturation of $\chi(\pi,\pi)$. Because of
the large $z$, extremely low temperatures are required and we have
therefore only gone up to $n = 192$, for which some clusters already
require $\beta = 2^{17}$ ($S(\pi,\pi)$ saturates at significantly lower
$\beta$ \cite{andersPRB2002}). The computational effort (CPU time and
memory) scales as $n\beta$ \cite{andersPRB1999}.

\begin{figure}
\includegraphics[width=6.75cm]{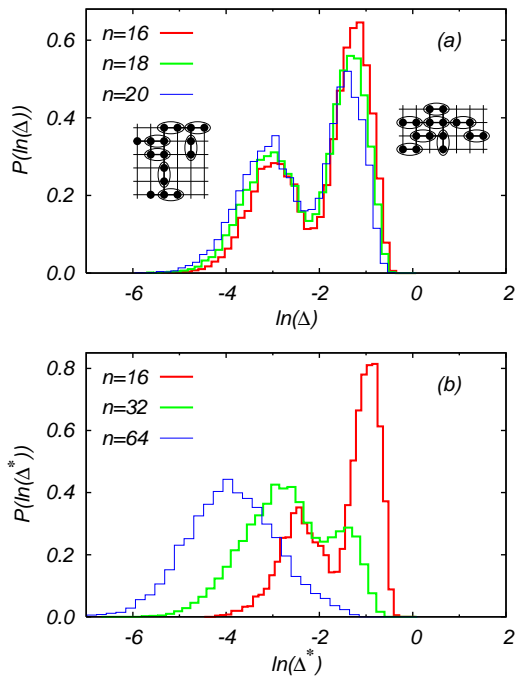}
\caption{(color online) (a) Singlet-triplet gap distribution for clusters  of 
size $n=16,18,20$. Representative clusters  corresponding to the 
two peaks are shown. (b) Distribution of the upper bound $\Delta^*$ for 
the gap for $n=16,32,64$.}
\label{lanczos-sse}
\end{figure}

In Fig.~\ref{lanczos-sse} we show the probability distributions of the
exact gap $\Delta$ of clusters with $n=16,18$, and $20$ obtained on the basis
of $4 \times 10^4$ samples for each size. We also show results for the
upper bound $\Delta^*$ for $n=16,32$, and $64$ obtained from
$6 \times 10^3$ clusters each. A common feature for the small clusters
is a double-peaked gap distribution. In the upper panel we show two
illustrative clusters corresponding to the two peaks. Identifying pairs
of spins likely to form singlets (giving a rough singlet-product approximation 
of the ground state) we find that the clusters of the lower peak always have
``dangling spins" not belonging to a local singlet (for any reasonable pairing; 
the locations of the dangling spins are of course not unique), whereas dangling 
spins are absent for the upper peak. The large-gap peak vanishes as $n$ grows; 
clearly because it is unlikely for a large cluster not to have any dangling spins.
Dangling spins arise from local imbalance in sublattice occupation. They should 
act as weakly interacting moments, with effective couplings decreasing with 
separation. One should hence expect small gaps for clusters with dangling 
spins at large separation. 

The role of isolated spins in the formation of long-range order has been 
pointed out by Bray-Ali and Moore \cite{moorePRB2004}. Although some 
spins can be very weakly coupled (effectively) to the rest of the cluster, 
correlations between them can be stronger than within the backbone of the 
cluster. Arbitrarily weakly coupled moments formed by groups of spins can 
also correlate over long distances and hence even a ``sloppy" fractal 
cluster can order at $T=0$ (a more thorough examination of long-range order was 
presented in \cite{moorePRB2006}). Here we will argue that excitations involving
localized effective moments are lower in energy than the quantum rotor states 
considered in previous discussions of the dynamic exponent \cite{vojtaPRL2005}.
 
\begin{figure}
\includegraphics[width=6.75cm]{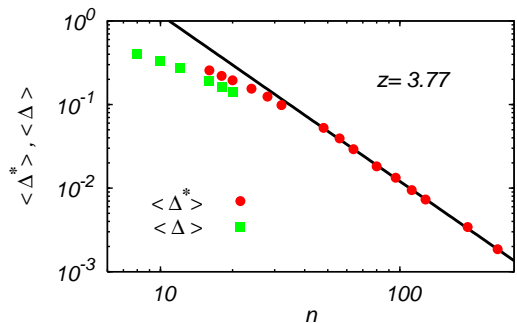}
\caption{(color online) Finite-size scaling of the average of the gap $\Delta$
and the upper bound $\Delta^*$. The slope of the line fitted to 
$\langle\Delta^*\rangle$ is $z/D_f = 1.99$.}
\label{z-scale}
\end{figure}

Fig.~\ref{z-scale} shows the size dependence of the
disorder averages $\langle \Delta\rangle$ and $\langle \Delta^*\rangle$.
A linear fit to $\langle \Delta^*\rangle$ gives a slope $z/D_f=1.99(2)$, 
or $z=3.77(4)$. Although the small systems for which we have 
calculated the exact gaps are not yet in the scaling regime, the available 
$\langle \Delta\rangle$ points reproduce well the shape of the $\langle 
\Delta^*\rangle$ curve in the range where they overlap. Comparing the $\Delta$ 
and $\Delta^*$ distributions for $n=16$ in Fig.~\ref{lanczos-sse}, we also see 
that they are very similar---$\Delta^*$ is only shifted up slightly 
relative to $\Delta$. 

\begin{figure}
\includegraphics[width=7.25cm]{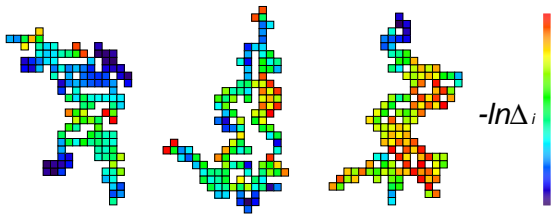}
\caption{(color online) Local gaps $\Delta_i$ for $n=128$ clusters, shown as
a color coded mapping of $-{\rm ln}(\Delta_i)$ to $[0,1]$ (for each 
cluster separately). The smallest gaps are  $\Delta_{\rm min}/J 
=0.027$ (left), $0.014$ (middle), and $0.012$ (right). The largest gaps are
$\Delta_{\rm max}/J = 0.9$ (left and right) and $0.22$ (middle).}
\label{local-excitation}
\end{figure}

The large $z$ implies that the lowest excitations of these clusters are not 
quantum rotor states (which would give $z=D_f \approx 1.89$). To investigate 
the nature of the excitations, we have calculated the
local susceptibility;
\begin{equation}
\chi_i=\int_0^{\beta}d\tau\langle S_i^z(\tau)S_i^z(0)\rangle,
\label{localx}
\end{equation}
which can be used to define a local gap \cite{rieger96}
$\Delta_i = 1/2\chi_i$, as in Eq.~(\ref{gapbound}). The magnitude 
of the local gaps of three representative clusters are illustrated in 
Fig.~\ref{local-excitation}. In the left cluster a few  isolated sites 
with small gaps can be distinguished, whereas the small gaps are more 
spread out in the one to the right. In the central one there are
isolated as well as more extended moments. Thus not only dangling spins but 
also other groups of spins can form effective moments. In all cases, the 
small-gap regions appear to be localized.  Regions with large gaps can 
be seen where it is apparent that the connectivity favors local 
singlet formation. Such singlets can contribute to isolating the
effective moments from each other.

Assuming that the low-energy physics of the clusters is governed by interactions 
between effective localized moments, we can apply concepts of the well-known 
strong-disorder renormalization group (RG) moment decimation procedure 
\cite{ma,fisher,bhatt} to analyze the low-energy behavior. The effective 
interactions between the moments can be ferromagnetic or antiferromagnetic, 
and we therefore have to consider two moments combining into either a smaller or a 
larger moment in each step \cite{westerberg} (including the case of two 
equal moments forming a singlet and dropping out). On a cluster with $n_A = n_B$ the 
decimation would ultimately leave two equal moments with an effective antiferromagnetic
interaction. The lowest excitation is the singlet-triplet excitation of this pair. 
A key question is whether the typical size of the renormalized moments grows with 
the cluster size or remains finite. If the moments remain finite the low-energy 
excitations are localized in the sense of involving a finite number of spins. 
However, the distance spanned by a pair of moments can grow with the cluster 
size, leading to decreasing effective interactions and thus smaller gaps.

\begin{figure}
\includegraphics[width=6.75cm, clip]{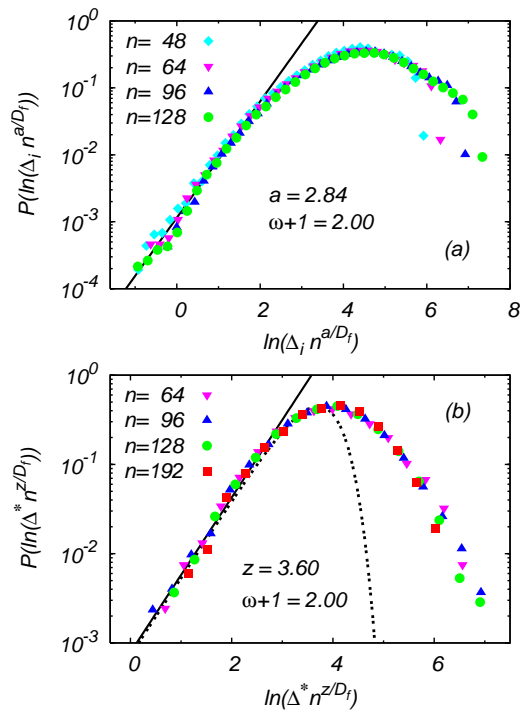}
\caption{(color online) (a) Distribution of the scaled gap 
$\epsilon_i=\Delta_iL^a$, with $a=2.84$. The line corresponds to a gap
exponent $\omega=1$. (b) Distribution of the upper bound $\Delta^*$ 
scaled with dynamic exponent $z=3.6$. The line shows the asymptotic small-gap 
behavior expected with $\omega=1$. The dashed curve is the 
Frechet distribution, Eq.~(\ref{u-distribution}).}
\label{gap-scaling}
\end{figure}

If the excitations indeed are localized moment-pair excitations, the 
distribution of the first gap (i.e., $\Delta$, which is here estimated as
$\Delta^*$ or the smallest of the local gaps, $\Delta_{\rm min}$) 
should be described by extremal value statistics \cite{extremebook,Lin} of the local gaps.
We thus consider the distribution of $\Delta_i$, taking into account a possible
scaling with the system size, as suggested by the discussion above. We define
$\epsilon_i = \Delta_iL^{a} \sim \Delta_in^{a/D_f}$. As shown in 
Fig.~\ref{gap-scaling}, for $a\approx 2.8$ the distributions of $\epsilon_i$
for different cluster sizes indeed collapse onto a single universal curve. For small 
$\epsilon_i$, $P(\epsilon_i)\sim A\epsilon_i^{\omega}$, where $\omega \approx 1$ 
and $A$ is a constant. We want the distribution $P_M(\epsilon_{\rm min})$ of 
the smallest local gap, 
$\epsilon_{\rm min} = \Delta_{\rm min}L^a={\rm  min}{\{\epsilon_1,\epsilon_2,\cdots,
\epsilon_M\}}$, for large $M$. If the excitations are localized we have 
$M \sim L^{D_f}$. Using the probability of a scaled local gap 
$\epsilon ' < \epsilon$,
\begin{equation}
 p_<(\epsilon)=\int_0^{\epsilon}P(\epsilon')d\epsilon'=
\frac{A}{\omega+1}\epsilon^{\omega+1},
\label{localgap-distribution}
\end{equation}
we find the typical $\epsilon_{\rm min}$ from $p_<(\epsilon_{\rm min}) \sim M^{-1}$.
Then using $M \sim L^{D_f}$ we get $\Delta_{\rm min} \sim L^{-a-D_f/(\omega+1)}$.
Since $\Delta_{\rm min}$ should scale as $\Delta$ (which we have also confirmed 
numerically); $\Delta_{\rm min} \sim L^{-z}$, we have
\begin{equation}
z=a+\frac{D_f}{\omega+1}.
\label{zrelation}
\end{equation}
This generalizes the relation $z=D_f/(\omega + 1)$ used as a criterion for a 
localized excitation by Lin et al.~\cite{Lin}.  With $a=2.84$ and $\omega=1$ 
from the universal $\Delta_i L^a$ curve in the upper panel 
of Fig.~\ref{gap-scaling}, we get $z=3.79$, in excellent agreement with the value 
$3.77(4)$ obtained in Fig.~\ref{z-scale} on the basis of $\langle \Delta^*\rangle$. 
The full probability distribution of $\Delta^*$ is also very consistent with 
$\omega=1$, as shown in the lower panel of Fig.~\ref{gap-scaling}. Here the best-fit 
dynamic exponent, $z \approx 3.6$, is slightly smaller than the value quoted above, 
but considering the limited number of cluster sizes and samples the agreement is 
quite satisfactory. We conclude that the scaling behavior is consistent with 
low-energy excitations due to finite local moment pairs.

We can also calculate the full distribution of $\Delta$. Using 
$P_M(\epsilon)=MP(\epsilon)[1-p_<(\epsilon)]^{M-1}$, or
\begin{equation}
 P_M(\epsilon) = -\frac{d}{d\epsilon}{\lbrack 1-p_<(\epsilon)\rbrack}^{M}
\simeq -\frac{d}{d\epsilon}e^{-Mp_<(\epsilon)}
\label{pm-distribution}
\end{equation}
and Eq.~(\ref{localgap-distribution}) gives the Frechet distribution 
\cite{extremebook},
\begin{equation}
P(u)=A{u}^{\omega}{\rm exp} (-A(\omega +1)^{-1}{u}^{\omega+1} ),
~~~u=u_0\Delta L^{z}.
\label{u-distribution}
\end{equation}
This form can only be fitted to the $\Delta^*$ data in Fig.~\ref{gap-scaling} 
in the small-gap region. It is not completely clear to us why it fails for 
larger gaps, but there are signs of slower size-convergence on this side of
the distribution. The deviations may also be related to the fact that we 
are studying the upper bound $\Delta^*$, not the exact gap $\Delta$. 

For clusters with $n_A\neq n_B$, the final result of the RG moment decimation 
procedure would not be two moments forming a singlet, but a single moment
with spin $S=|n_A- n_B|/2$. The global sublattice imbalance would necessarily lead 
to moments that on average grow as the decimation proceeds, unlike in the case of 
$n_A=n_B$ where it is possible that the effective moments reach a 
steady-state size as they are equally likely to grow or shrink at every RG step
(a steady-state size is of course not guaranteed on purely statistical grounds, 
but our scaling study above supports it). The lowest excitation for an $n_A\neq n_B$
cluster must hence involve spins scattered through the whole cluster. Considering 
that $\langle S\rangle \sim \sqrt{n}$, such an excitation cannot be local. We indeed 
find that the exponent relation (\ref{zrelation}) is violated for clusters 
with even $n$ but no further restrictions on $n_A$ and $n_B$. We get $a \approx 1.9$ 
from scaling $\Delta_i$, which is considerably smaller than $a \approx 2.8$ for 
$n_A=n_B$. However, the dynamic exponent obtained by scaling $\Delta^*$ still
remains $z \approx 2D_f$.

To conclude, we have presented a scaling study of the excitations 
of percolating $S=1/2$ Heisenberg clusters, showing that the dynamic exponent 
$z = 3.7(1)$; a factor 2 larger than $z=D_f = 91/48$ expected for quantum 
rotor states \cite{vojtaPRL2005}. We propose that the low-energy excitations 
are due to interactions between effective moments formed by local imbalance in 
sublattice occupation. We have used concepts of strong-disorder RG and extremal
value statistics to derive a scaling relation, Eq.~(\ref{zrelation}), between $z$ 
and two exponents governing the scaling of local gaps, under the condition that the 
low-energy excitations involve a finite number of spins. For clusters with an equal 
number of spins on the two sublattices ($n_A=n_B$), this relation is satisfied by 
numerical results for an upper bound of the gap, $\Delta^*$, defined in Eq.~(\ref{gapbound}), 
and local gaps $\Delta_i$ similarly defined. However, for clusters with $n_A\not= n_B$ 
the global sublattice imbalance implies that the renormalized moments grow with 
system size. In accord with this non-locality we find that our exponent 
relation is violated. However, the dynamic exponent $z \approx 2D_f$ for both
$n_A = n_B$ and $n_A\not= n_B$ clusters.

Our scenario implies that once two identical clusters are coupled in a 
bilayer, with a small inter-layer coupling $J_\perp$ (smaller than the value 
at which the long-range order vanishes \cite{awsbilayer}), the low-energy excitations 
should change as effective moments in opposite layers comletely compensate each other. 
Preliminary calculations for clusters with $J_\perp/J \approx 0.05$ indeed show a much 
smaller $z$, consistent with $z=D_f$ and the quantum rotor picture \cite{vojtaPRL2005}. 
This is also the case for the 3D classical Heisenberg model with columnar defects 
\cite{vojtacondmat}, which should describe the bilayer but not the single layer for 
which there are uncompensated Berry phases (local sublattice imbalance) not 
captured by the mapping to the classical model.

Another type of low-energy excitation possible in a percolating cluster is a 
fracton \cite{orbachModernPhy1994,antonioPRB2004}. The fracton is localized for 
finite energy $\Omega$, but delocalizes as $\Omega \to 0$. Hence, based on our 
scaling results for $n_A=n_B$, it is different from our effective moment excitation. 
For a finite system $\Omega(L) \propto L^{-D_f/d_f}$, where $d_f$ is the fracton 
dimensionality. Approximate calculations of $d_f$ \cite{orbachModernPhy1994} 
have given $z_f = D_f/d_f$ smaller than the $z$ reported here.

We would like to thank A.~Castro Neto, C.~Henley, J. Moore, S. Sachdev, J.~Schmalian, 
and T. Vojta for  useful discussions and comments.  This work is supported by the 
NSF under Grant No.~DMR-0513930.

\null

\end{document}